\documentclass[aps,prr,twocolumn,groupedaddress]{revtex4-2}
\usepackage{color}

\usepackage[utf8]{inputenc}
\usepackage[english]{babel}
\usepackage{amsmath,graphicx,enumerate}

\begin{document}

%\preprint{}

%\title{Activity-induced mechanism for clogging of micro-channels.} 
%\title{Active-particles-based clogging mechanism} 
\title{Activity-controlled clogging and unclogging of micro-channels} 

\author{L. Caprini$^{1}$}
\author{F. Cecconi$^{2}$}
\author{C. Maggi$^{2}$} 
\author{U. Marini Bettolo Marconi$^{3}$}
\affiliation{$^1$ Gran Sasso Science Institute (GSSI), Via. F. Crispi 7, 67100 L'Aquila, Italy.\\ 
$^2$ Istituto dei Sistemi Complessi - CNR and Dipartimento di Fisica, Universit\`a di Roma Sapienza, P.le Aldo Moro 2, 00185, Rome, Italy\\
$^3$ Scuola di Scienze e Tecnologie, Universit\`a di Camerino - via Madonna delle Carceri, 62032, Camerino, Italy.\\ 
 }

\date{\today}

\begin{abstract}
We propose a mechanism to control the formation of stable obstructions in two-dimensional micro-channels of variable sections taking advantage of the peculiar clustering property of active systems. 
Under the activation of the self-propulsion by external stimuli, the system behaves as a switch according to the following principle:
by turning-on the self-propulsion the particles become active and even at very low densities stick to the walls and form growing layers eventually blocking the channel bottleneck, while 
the obstruction dissolves when the self-propulsion is turned off.
We construct the phase diagram distinguishing clogged and open states in terms of density and bottleneck width. 
The study of the average clogging time, as a function of density and bottleneck width, reveals the marked efficiency of the active clogging that swiftly responds to the self-propulsion turning on.
The resulting picture shows a profound difference with respect to the clogging obtained through the slow diffusive dynamics of attractive passive Brownian disks.
This numerical work suggests a novel method to use particles with externally tunable self-propulsion to create or destroy plugs in micro-channels.
\end{abstract}

\maketitle

%%%%%%%%%%%%%%%%%%%%%%%%%%%%%%%%%%%%%%%%%%%%%%%%%%%%%%%%%%%%%%%%%%%%%%%%%%%
%\section*{INTRODUCTION}
%%%%%%%%%%%%%%%%%%%%%%%%%%%%%%%%%%%%%%%%%%%%%%%%%%%%%%%%%%%%%%%%%%%%%%%%%%%
%Structure of the Introduction

%\begin{itemize}

%\item[1)] Present the clogging (usually something to avoid) and Illustrate the possible applications: control the plugs formation (clogging/unclogging)

%\item[2)] Introduction to active particles: focus on artificial microswimmers bacteria etc, and mostly on microswimmers sensible to ligth (or which can be activated by external imput).  

%\item[3)] Formation of clusters, MIPS, etc also experimental works (focus on clustering induced by ligths)

%\item[4)] Clustering in the proximity of obstacles (also pillars, funnels), rigid boundaries etc (the cluster formation is enhanced).

%\item[5)] The idea is to use the clogging induced by the clustering of active particles to close or open a canal.

%\end{itemize}

\section{Introduction}

Several technological and industrial processes require the control of fluid flows through channels and pores at mesoscopic scales. In this context, it is important to find strategies either favoring or preventing the sudden blockage (clogging) of the channels by particle aggregates and cohesive matter~\cite{dressaire2017clogging}.
Recently, materials that spontaneously respond to environmental changes, known as smart materials, seem to offer new opportunities for a clever solution to this kind of problem.
Smart materials can also be used to deliver cohesive substances into specific regions in order to reinforce surfaces and repair fractures or damages~\cite{van2009simulations, zhang2012review}. 
In principle, the material aggregation could be used to form obstructions capable of blocking the passage of undesirable debris or harmful chemical and biological agents. 
%%%
In this Letter, we provide a proof of concept that self-propelled particles~\cite{marchetti2013hydrodynamics,bechinger2016active, berthier2019glassy,gompper20202020}, whose active force can be controlled by external inputs~\cite{palacci2013living}, can be employed 
as smart materials~\cite{vernerey2019biological} able to generate removable obstructions into channels by aggregation.
Indeed, it has been recently shown that genetically engineered \textit{E.~Coli} bacteria~\cite{walter2007light,arlt2018painting,frangipane2018dynamic} and certain Janus particles~\cite{palacci2013living,buttinoni2012active,volpe2011microswimmers, schmidt2019light} can be externally controlled by a light stimulus and their activity can be rapidly switched on/off by modulating the illumination power that could be employed to design active
rectification devices~\cite{stenhammar2016light}.
%{\color{red} At variance with Ref.~\ref{yu2016confined}, where a single Janus particle is used to push aggregates of passive particles out of a channel, here,} 
Specifically,
we suggest taking advantage of the self-propelled particle propensity to spontaneously form stable aggregates and undergo motility induced phase separation 
(MIPS)~\cite{cates2015motility, bialke2015active, ma2020dynamic, fily2012athermal, gonnella2015motility},
as experimentally observed for artificial microswimmers~\cite{buttinoni2013dynamical, palacci2013living, bialke2015active, ginot2018aggregation, Dhruv2017, van2019interrupted} 
or bacteria~\cite{peruani2012collective, petroff2015fast, dell2018growing} and reproduced by numerical simulations~\cite{caprini2020spontaneous, caprini2020hidden, bialke2013microscopic, speck2016collective, liebchen2017collective, levis2017active, solon2015phase, tjhung2018cluster, chiarantoni2020work, jose2020phase, mandal2019motility, shi2020self}. 

Our mechanism based on the clustering of active particles is able to work as a switch to clog/unclog channels by turning on/off the self-propulsion. Its usefulness is also suggested by the low particle concentrations required. In fact, the cluster formation is strongly enhanced by the presence of confining geometries, as experimentally shown for bacteria~\cite{costanzo2012transport, figueroa2015living, yawata2016microfluidic} or artificial microswimmers~\cite{simmchen2016topographical, volpe2011microswimmers}, since active particles accumulate near boundaries~\cite{caprini2018active, maggi2015multidimensional, wittmann2016active, das2020aggregation}, wall channels~\cite{wensink2008aggregation, khodygo2019homogeneous, caprini2019active, yang2014aggregation, kudrolli2008swarming}, and obstacles~\cite{harder2014role, ray2014casimir, ni2015tunable, knevzevic2020capillary}.
The employment of active particles, instead of passive colloidal particles, to control the channel occlusion leads to further advantages.
Passive particles cluster only in the presence of attractive interactions and the addition of depletants~\cite{kilfoil2003dynamics,hobbie1998metastability} but, even in these cases, exhibit a very slow dynamics.
As a consequence, the clogging formation is very slow and has been experimentally and numerically observed only when accelerating the access of passive particles into the channel, for instance, by imposing external fluid flows~\cite{dersoir2017clogging, sauret2018growth, cejas2018universal, pal2019quantitative}. 
By contrast, as we shall see, active particles block the channel in a much shorter time than passive colloids, revealing their prominent efficiency. 
This property is crucial in view of the possibility of achieving efficient switching-like behaviors to clog/unclog channels.
We outline that the mechanism presented here differs from the experimental study of Ref.~\cite{yu2016confined} where a single active colloid is used to push aggregates of passive particles out of a channel using the persistence of the active dynamics.

%{\color{red} To the best of our knowledge, previous studies on active particles in channel just focus on other aspect~\ref{yu2016confined}}

The article is structured as follows: In Sec.~\ref{Sec:model}, we describe the model numerically studied, in particular, the geometrical set-up and the equation of motion of the active particles. 
The numerical and theoretical results are reported in Sec.~\ref{Sec:results} while discussions and conclusions are presented in the final section.

%------------------------ FIG.1--------------------------------------
\begin{figure}[!t]
\centering
\includegraphics[width=1\linewidth,keepaspectratio]{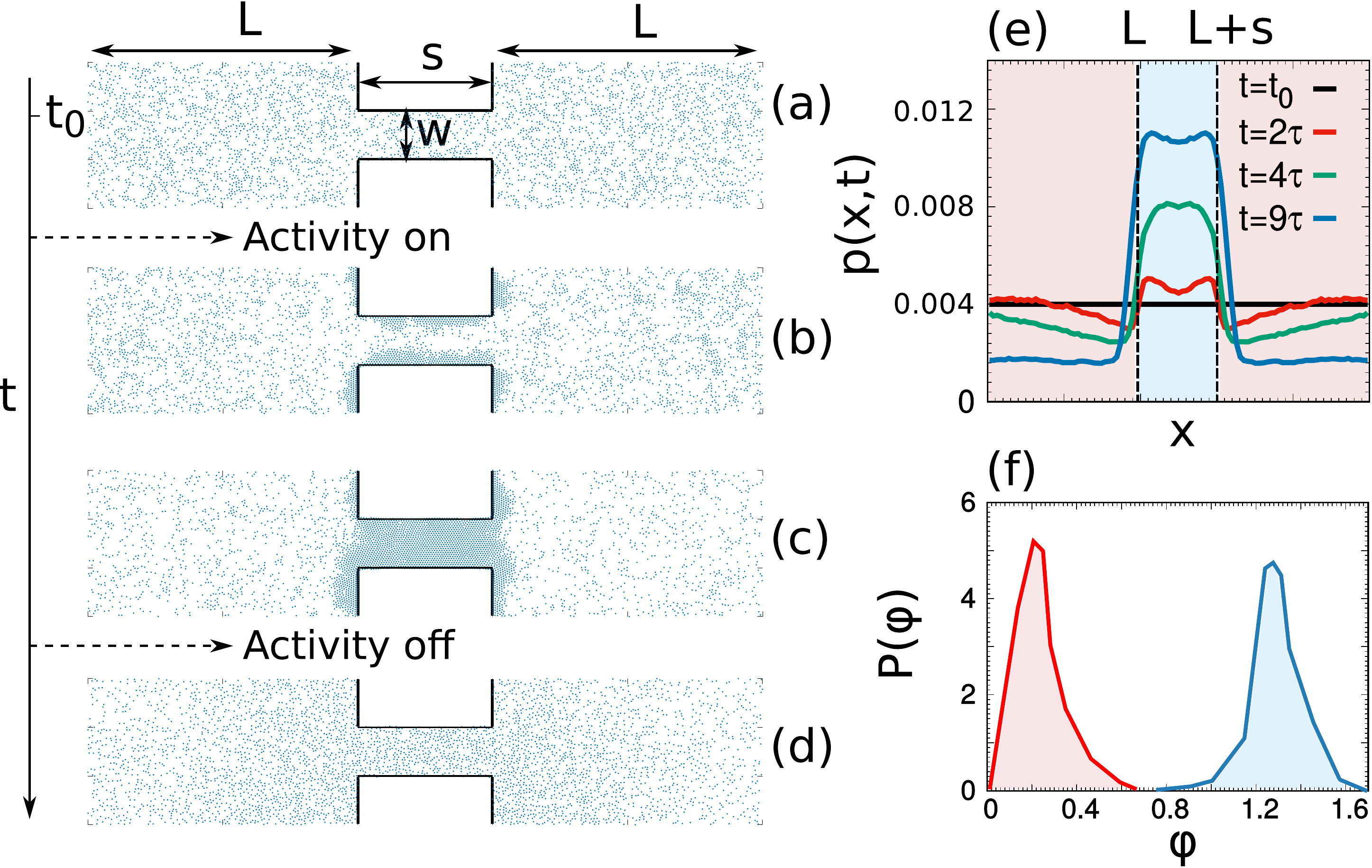}
\caption{\label{fig:picture}
Phenomenology of the clogging process. 
Panels~(a)-(d) show the snapshots of the typical time evolution of the system, with $w=20$, $\rho=0.3$, $D_r=1$, $v_0=25$. 
In particular, panel~(a) reports the starting homogeneous configuration in the absence of self-propulsion. 
Panel~(b) and~(c) are obtained before and after the plug formation after the active force is turned on.
Finally, in panel~(d), we report a configuration relaxing towards the homogeneous state since the active force is turned off. 
In panel~(e), we study the particle density, $p(x, t)$, for different times, where
the light blue and the pink regions mark the bottleneck and the lateral boxes, respectively. 
Panel~(f) shows the stationary distribution of the packing fraction, $P(\phi)$, calculated in the bottleneck (blue curve) and in the lateral boxes (red curve) for the parameter setting of the other panels.}
\end{figure}
%---------------------------------------------------------------------

\section{Model}\label{Sec:model}

We consider a system of $N$ interacting self-propelled disks in two dimensions.
According to the Active Brownian Particles (ABP) dynamics \cite{fodor2018statistical, shaebani2020computational}, the self-propulsion is modeled as a time-dependent 
force with fixed modulus, $v_0$, and orientation, $\mathbf{n}_i=(\cos{\theta_i}, \sin{\theta_i})$, 
where the angles, $\theta_i$, evolve as independent Wiener processes. 
The dynamics of the particle positions, $\mathbf{x}_i$, is governed by overdamped stochastic equations:
\begin{subequations}
\label{eq:wholeABPdynamics}
\begin{align}
\label{eq:xf_dynamics}
\gamma\dot{\mathbf{x}}_i &= \mathbf{F}_i  + \mathbf{F}^w_i + \gamma v_0 \mathbf{n}_i  \\
\label{eq:theta_dynamics}
\dot{\theta}_i&= \sqrt{2D_r} \xi_i  \,,
\end{align}
\end{subequations}
where $\xi_i$ is a white noise with unit variance and zero average.
The constants $\gamma$ and $D_r$ denote the friction and the rotational diffusion coefficients, respectively, and, in particular, the latter determines the typical persistence time of the active force, $\tau=1/D_r$.
The first force term, $\mathbf{F}_i=-\nabla_i U_{tot}$, models the steric repulsion between two disks, where
$U_{tot} = \sum_{i<j} U(|{\mathbf r}_{ij}|)$ with ${\mathbf r}_{ij}= \mathbf{x}_i -\mathbf{x}_j$ and
the shape of $U$ is chosen as a truncated and shifted Lennard-Jones potential,
$U(r)=4\epsilon[(\sigma/r)^{12} - (\sigma/r)^{6}] + \epsilon$ for $r\leq 2^{1/6}\sigma$ and zero otherwise. The constants $\sigma$ and $\epsilon$ represent the particle diameter and the energy scale of the interactions, respectively.
%The term, $\mathbf{F}^w_i$, is the repulsion exerted by  the rigid boundaries, modeled as soft reflecting walls, derived by the potential, $V(k(x)-y)$, where {\color{red} $V$ is a repulsive, truncated harmonic potential} and the curve $y=k(x)$ is the wall profile.
%
%being $\hat{\mathbf{n}}$ the normal direction with respect to $y=k(x)$, such that any tangential or torque contributions exerted by the boundaries are neglected.
The term, $\mathbf{F}^w_i$, is the repulsion force exerted on the active particles by the boundaries defined by the profile $y=k(x)$. 
Each boundary repels the particles crossing the curve $y=k(x)$ outward
with a stiff harmonic force that re-injects them inside the channel.
This force is derived by a stiff truncated harmonic potential,
$$
W(x,y) = \dfrac{A}{2}[k(x)-y]^2 \Theta[k[x]-y]
$$
where $\Theta$ is the unitary step function, and $A=10^3$ is the strength of the potential chosen to ensure the impenetrability of the channel walls.
The force exerted by the walls can be calculated as $\mathbf{F}^w_i=-\nabla W(x,y)$ that is proportional to the normal $\hat{\mathbf{n}}$ with respect to the curve $y=k(x)$, such that any tangential contribution from the boundaries is neglected.
Further details about the implementation of the boundaries (and the functions $k(x)$) are reported in Appendix~\ref{Sec:Sec1}.
In practice, the system consists of two boxes of area $L \times H$ connected by a bottleneck of size $s \times w$, as schematically illustrated in Fig.~\ref{fig:picture}~(a).
In particular, the solid black lines 
define the bottleneck region, while periodic boundary conditions 
are applied to the rest of the channel.
%---------------------------- FIG.2 --------------------------------
\begin{figure}[!t]
\centering
\includegraphics[width=0.9\linewidth,keepaspectratio]{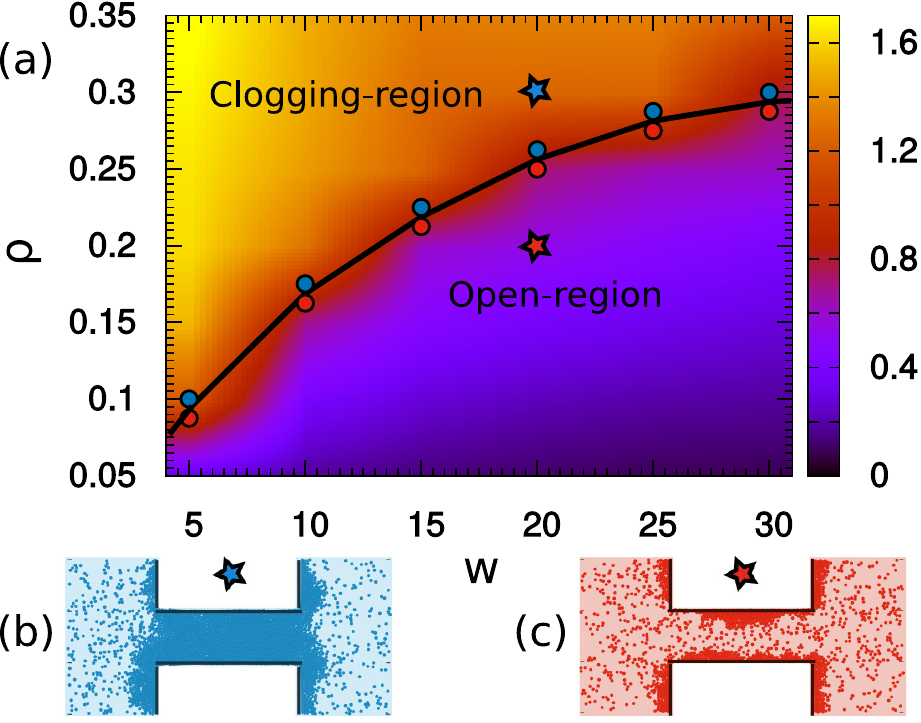}
\caption{\label{fig:phasediag}
Clogging phase diagram. Panel~(a): phase diagram as a function of bottleneck width, $w$, and density of the system, $\rho$.
Colors represent the steady-state density values in the bottleneck, $\langle \rho_b\rangle$.
The solid black curve indicates the clogging-line numerically obtained, separating clogged from open states. 
Blue and red circles mark the lowest and the highest values of $\rho$ at which clogged and open states are still observed.
Panels~(b) and~(c) report two zooms of the bottleneck occupation above and below the clogging line, in correspondence of the colored stars in the phase diagram.
Simulations are run with $D_r=1$, $v_0=25$, $L=100$, $H=60$ and $s=50$. 
}  
\end{figure}
%---------------------------------------------------------------------

%%%%%%%%%%%%%%%%%%%%%%%%%%%%%%%%%%%%%%%%%%%%%%%%%%%%%%%%%%%%%%%%%%%%%
\section{Results}\label{Sec:results}
%%%%%%%%%%%%%%%%%%%%%%%%%%%%%%%%%%%%%%%%%%%%%%%%%%%%%%%%%%%%%%%%%%%%%
\subsection{Activity-induced clogging}
We study the dynamics~\eqref{eq:wholeABPdynamics} at fixed active force by varying the density, $\rho$, and the bottleneck width, $w$. 
Simulations started from homogeneous configurations, as shown in Fig.~\ref{fig:picture}~(a), that are the typical spatial configurations of passive colloids before the particle activation.
At the initial time $t_0$, we turn on the self-propulsion and let the system evolve for a final time, $T =10^2 /D_r$.
A typical time evolution, at $\rho=0.3$ and $w=20$, 
is schematically illustrated in panels (b), (c) and (d).
In a first transient regime, particles accumulate in front of the bottleneck walls forming two symmetric growing layers, as illustrated in Fig.~\ref{fig:picture}~(b).
Subsequently, the two layers coalesce and clog the bottleneck, 
Fig.~\ref{fig:picture}~(c), forming a very dense and cohesive cluster as revealed by the bimodal shape of the density distribution, plotted in Fig.~\ref{fig:picture}~(f).
Finally, the plug dissolves when the active force is switched off and the system gradually recovers a homogeneous configuration, Fig.~\ref{fig:picture}~(d).
In this case, the dynamics is purely diffusive and controlled by the translational diffusion coefficient that is no more negligible after the activity turning off. 
Despite the intrinsic slowness of the diffusive dynamics, the cohesive plug dissolves rapidly since the clogged configurations are very far from the equilibrium configurations of passive systems.
To accelerate the depletion of the clogged region, we can raise the temperature to increase the diffusivity. Alternatively, we can use the spatial-modulation of the light to induce inhomogeneity that destabilizes the cluster cohesion.
The scenario described by panels (a)-(d) is quantitatively confirmed in Fig.~\ref{fig:picture}~(e) where the time-dependent density distribution, $p(x,t)$, along the channel, is plotted at different times.
Thus, by turning the self-propulsion on/off, the system, in practice, behaves as a switch to clog and unclog the channel.
However, for $w$ or $\rho$ values small enough,
the system is not able to attain a steady-state with stable 
bottleneck obstructions as reported in Fig.~\ref{fig:phasediag}~(c) where the steady-state is characterized by
small clusters of particles close to the bottleneck walls (that will be denoted as ``open state'' along with the rest of the paper).

The distinction between clogged and open states can be achieved by computing the average density in the bottleneck region, $\langle \rho_b \rangle$, after the $\rho$-trajectories have reached their plateau. 
A close inspection of the configurations allows us to verify that clogged states are those with $\langle \rho_b \rangle\gtrsim 1$, while open states
correspond to $\langle \rho_b \rangle\lesssim 1$. 
Through this heuristic criterion, we construct the phase diagram 
of the system as a function of $\rho$ and $w$, Fig.~\ref{fig:phasediag}~(a), where clogged and open configurations are separated by a solid black line (clogging-line).
The clogging-line displays a monotonic growth with both 
$\rho$ and $w$ almost saturating around $\rho=0.3$, which is well below the critical $\rho$-value to observe the MIPS-transition in the confinement-free system~\cite{redner2013structure, solon2018generalized, digregorio2018full, petrelli2018active}.
The color-map encodes the values of $\langle \rho_b \rangle$ in the bottleneck region showing that a sharp color variation occurs in the proximity of the clogging line and, in the clogged states, plugs become less cohesive as $w$ grows. 
This phase diagram indicates the working operational conditions of the ``switching device'' at different channel widths also showing that the clogged states are obtained even at very small densities. 
The low-density working condition constitutes a strong advantage in the potential applicability of our mechanism to real devices.

%We remark that the clogged states are obtained even at very small densities constituting a strong advantage in the potential applicability of our mechanism to real devices.

%------------------------------- FIG.3 ------------------------------------------
\begin{figure}[!t]
\centering
\includegraphics[width=1\linewidth,keepaspectratio]{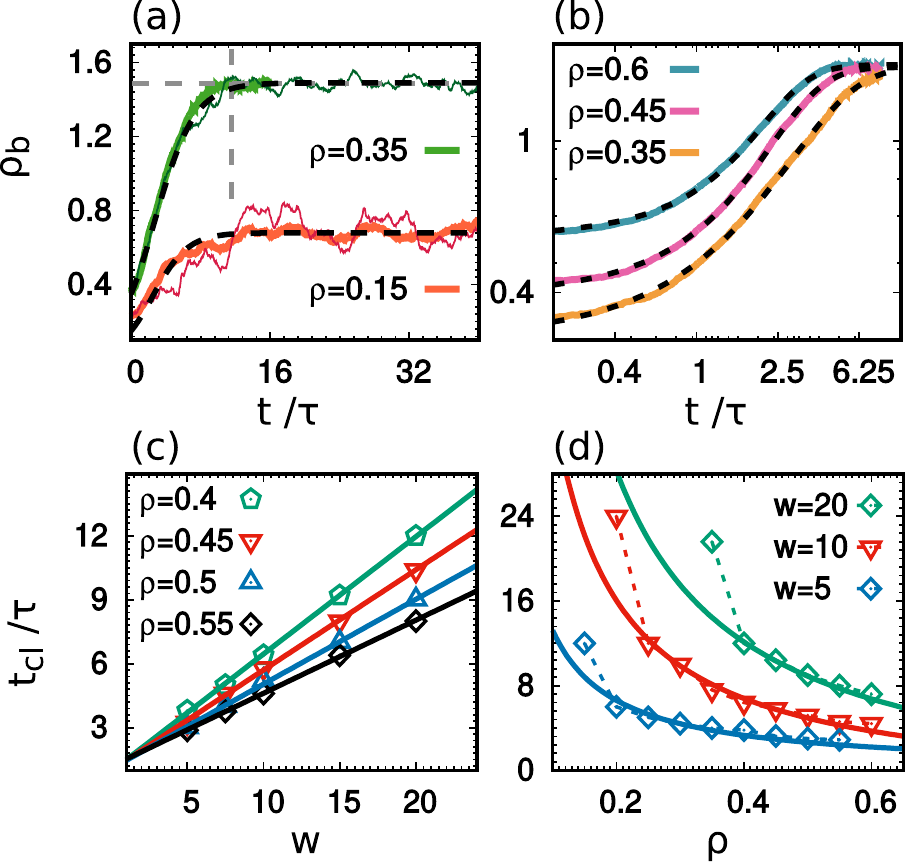}
\caption{\label{fig:timeandtraj}
Dynamics of the plug formation.
Panels~(a) and~(b) plot the time evolution of the density in the bottleneck, $\rho_b(t)$, for $w=10$. 
Panel~(a) shows single-trajectories (thin lines) and the ensemble average,
$\langle\rho_b(t)\rangle$ (tick lines), taken over 20 independent initial configurations. 
The dashed grey horizontal line marks the asymptotic value $\langle\rho_b\rangle$, while the dashed vertical line indicates the time at which the asymptotics is reached.
Panel~(b) plots $\langle\rho_b\rangle$ as a function of $t/\tau$ for three different values of $\rho$ leading to clogging configurations.
Both in panels~(a) and~(b), dashed black lines are obtained by a fit to data with Eq.~\eqref{eq:rhob(t)}.
Panel (c) contains the clogging time $t_{\text{cl}}$ as a function of $w$ for different values of $\rho$, where solid lines are obtained from numerical linear fits.
Panel (d) reports $t_{\text{cl}}$ vs $\rho$ for different values of $w$. 
Points are obtained from simulations, solid lines from Eq.~\eqref{eq:prediction_tw} and dotted lines are just eye-guides.
Simulations are performed with $\tau=1/D_r=1$, $v_0=25$, $L=100$, $H=60$ and $s=50$. 
}  
\end{figure}
%--------------------------------------------------------------------------------------------

\subsection{The dynamics of the active clogging}
To work as a switching mechanism, the clogging process needs to be 
sufficiently swift in the response to the turning on of the active force. 
In this respect, we monitor the time behavior of the local density in the bottleneck, $\rho_b(t)$.
Figs.~\ref{fig:timeandtraj}~(a)-(b) illustrate the typical dynamics of the clogging process for a bottleneck of width $w=10$.
In particular, panel (a) compares the single fluctuating trajectory of $\rho_b(t)$ with its ensemble average $\langle \rho_b(t) \rangle$ for two different values of $\rho$. 
All the curves saturate at a plateau whose value indicates the clogging degree of the stationary state. 
Specifically, the higher and lower values correspond to clogged (green curves) and open states (red curves), respectively.
%In addition, the former are self-averaging while the latter are still characterized by more fluctuating behaviors, even in the steady-state configurations, since the layers of particles attached to the walls reorganize without merging.
In addition, in the the former case, $\rho_b(t)$ displays very small fluctuations around $\langle \rho_b(t) \rangle$ while, in the latter case, the $\rho_b(t)$ shows larger fluctuations, even in the steady-state configurations, since the layers of particles attached to the walls reorganize without merging.
The dashed lines in Fig.~\eqref{fig:timeandtraj}~(a) represent
the theoretical predictions of $\langle\rho_b (t) \rangle$ 
\begin{equation}
\label{eq:rhob(t)}
\langle\rho_b(t)\rangle = \frac{\langle\rho_b\rangle \rho}{\rho + \left(\langle\rho_b\rangle - \rho\right) e^{-t/\alpha}} \,, 
\end{equation}
where $\alpha$ is a fitting parameter.
Eq.~\eqref{eq:rhob(t)} is the solution of the logistic equation which, for the present system, is derived in Appendix~\ref{Sec:Sec2} under suitable approximations, observing that the increase of $\langle\rho_b(t)\rangle$ is mainly determined by the particles approaching almost \emph{ballistically} the bottleneck and that the probability to remain trapped is roughly proportional to $\rho_b(t)$. 
The prediction~\eqref{eq:rhob(t)} reveals also a good agreement with the numerical results for $\langle \rho_b(t) \rangle$
as shown in Fig.~\ref{fig:timeandtraj}~(b) for several values of $\rho$ giving rise to clogged configurations. 
In these cases, $\langle\rho_b(t)\rangle$ saturates at a common plateau that is determined by the maximum packing density in the bottleneck.

The temporal delay of the switch can be estimated as the time, $t_{\text{cl}}$, needed to observe the plug formation in the channel (clogging time).
Operatively, $t_{\text{cl}}$ is measured as the time such that $\langle \rho_b(t)\rangle$ attains the asymptotic value $\langle \rho_b\rangle$ with an uncertainty of 5\%.
To characterize the switching efficiency, we study the dependence of $t_{\text{cl}}$ on the bottleneck width $w$ and density $\rho$.
Fig.~\ref{fig:timeandtraj}~(c) shows the linear scaling of $t_{\text{cl}}$ as a function of $w$ for different values of $\rho$ (straight lines are linear fits to data).
Instead, Fig.~\ref{fig:timeandtraj}~(d) reports the monotonic decrease of $t_{\text{cl}}$ with $\rho$, showing that for low values of $\rho$ the onset of the clogging state is prohibitive in time. However, in view of the possible applications, it is encouraging that there exists an extensive range of $w$ and $\rho$ where $t_{\text{cl}}$ is only of the order of a few persistence times, $\tau$, of the self-propulsion. 
An analytical prediction of $t_{\text{cl}}$ can be obtained
upon the assumption that the plug formation very weakly affects the bulk average density (large lateral boxes): 
\begin{equation}
\label{eq:prediction_tw}
t_{\text{cl}} \approx \frac{s w}{\mathcal{R}}\frac{D_r}{2 v_0^2} \left(\frac{\left\langle\rho_b\right\rangle}{\rho} -1\right) \,,
\end{equation}
where $\mathcal{R}$ is a geometrical factor. More details about the derivation of the prediction~\eqref{eq:prediction_tw} are reported in Appendix~\ref{Sec:Sec3}.
The comparison with data in Fig.~\ref{fig:timeandtraj}~(d) reveals a good agreement except for the range of low values of $\rho$, where Eq.~\eqref{eq:prediction_tw} underestimates $t_{\text{cl}}$ because the hypothesis of almost constant bulk-density is no 
longer applicable.

The above results are very promising from a practical perspective to design real switching devices based on the active clogging. One can argue that the same process could be obtained 
through the coarsening of passive attractive colloids upon the introduction of wall-attractive interactions via chemical coating of the bottleneck walls.
This possibility can be tested by replacing active with attractive passive particles in the presence of attractive bottleneck walls.
The details of the passive numerical study and the corresponding results are discussed in Appendix~\ref{Sec:Sec4}.
However, our simulations do not show any bottleneck obstruction within the typical times taken by the active system to approach the clogged state.
Indeed, the simple self-diffusion alone constitutes a very slow transport mechanism, as supported by direct simulations reported in Appendix~\ref{Sec:Sec6}, where a system of independent passive particles escape a box across two lateral holes, mimicking the presence of the bottleneck. 
To get a qualitative idea of the $t_{\text{cl}}$-scaling with the bottleneck width in the passive clogging process, we resort to Monte Carlo simulations of an equilibrium attractive lattice gas within the channel considered so far.
Appendix~\ref{Sec:Sec5} shows the scaling $t_w \propto w^2$ independently of the temperature that, in comparison with the linear scaling of the active $t_{\text{cl}}$, corroborates the idea that the formation of plugs in passive systems is less efficient.
As a conclusion, passive colloids cannot be considered as good candidates for the implementation of switches similar to those suggested in this work.

%-------------------------- FIG.4 ------------------------------------------------------
\begin{figure}[!t]
\centering
\includegraphics[width=1\linewidth,keepaspectratio]{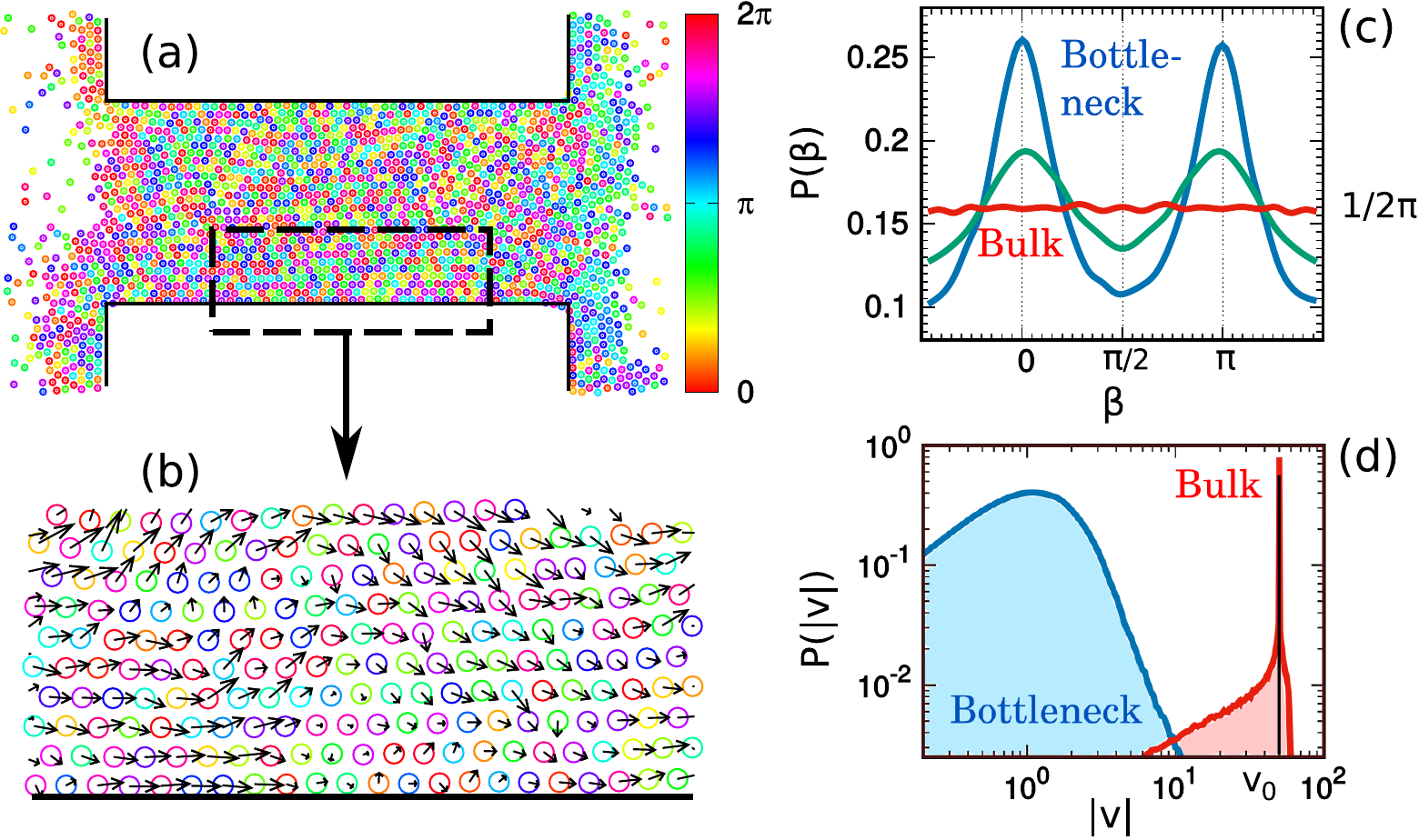}
\caption{\label{fig:Velocity}
Panel~(a) illustrates a snapshot configuration realized with $w=20$ and $\rho=0.3$ at time 
$T=50 \tau$, while panel~(b) is a zoom of the area delimited by the dashed black line in panel~(a). 
Colors encode the orientations of the particle self-propulsion with respect to the $\hat{\mathbf{x}}$-axis, while black arrows represent the velocity vectors.
Panel~(c) shows $P(\beta)$, the steady-state distributions of the velocity orientation. 
The blue distribution is in a small layer of width $H/5$ attached to the walls while the green one in a layer of width $H/5$ placed at the center of the bottleneck. 
Finally, the red $P(\beta)$ is given by averaging the velocity in a square region in the bulk of the lateral boxes. 
In panel~(d), we report $P(|\mathbf{v}|)$, i.e. the steady-state distribution of the velocity modulus, where blue and red curves are obtained averaging the velocities in the bottleneck and bulk regions of the lateral boxes.
Simulations are run with $D_r=1$, $v_0=25$, $L=100$, $H=60$ and $s=50$.}  
\end{figure}
%------------------------------------------------------------------------------------------------

%===============================================================================
\subsection{Steady-state properties of the plug}
%================================================================================
%{\color{blue} Introduce this study in the proper manner.}\\
To fully understand the active clogging mechanism, 
we also need to characterize the steady-state properties of the plug in a typical clogged configuration and, in particular, its dynamical properties.
Future investigations will aim to address the question of the stability of the clogging mechanism.

%To understand the dynamical properties of the plug, 
We study the typical configuration of a clogged bottleneck, 
that is reported in Fig.~\ref{fig:Velocity}~(a) where
the particles are colored according to the orientations of their self-propulsion, $\theta_i$.
Since the angles $\theta_i$'s evolve independently (see Eq.~\eqref{eq:theta_dynamics}), 
colors are randomly distributed in the whole system. %both in the lateral boxes and in the bottleneck.
However, the particle velocities, $\mathbf{v}_i=\dot{\mathbf{x}}_i$, tend to spontaneously align with each other, revealing the emergence of large aligned domains~\cite{caprini2020spontaneous, caprini2020hidden, caprini2020time}, whose particles have a common velocity orientation, $\beta_i$, with respect to the $\hat{\mathbf{x}}$-axis (Fig.~\ref{fig:Velocity}~(b)).
Near the bottleneck boundaries, the velocity orientations become preferentially parallel to the walls, as revealed by the symmetric peaks in $(0,\pi)$ of the steady-state distribution $P(\beta)$, Fig.~\ref{fig:Velocity}~(c). 
Moving towards the middle of the bottleneck, the peaks broaden 
as shown in Fig.~\ref{fig:Velocity}~(c) for two sections placed at the wall and the middle of the bottleneck
(for comparison we report also the $P(\beta)$ in the bulk of the lateral boxes, which 
is completely flat due to the absence of preferential orientations).
Fig.~\ref{fig:Velocity}~(d) compares the distribution of the single-particle velocity modulus in the bottleneck and lateral boxes.
In the latter case, the distribution is peaked around $v_0$, coinciding with the velocity modulus of a free independent self-propelled particle.
Instead, in the bottleneck, the distribution is peaked at a value of $|\mathbf{v}|\ll v_0$.

As a conclusion of this section, we remark that the formation of velocity aligned domains could, in principle, suggest the hindering of the plug stability (with the creation of fracture lines), while the slow particle motion in any clustered configuration should play the opposite role.
In future studies, we will check the stability of the mechanism proposed in this paper, testing if the activity-induced obstruction is able to really block the passage of large colloidal tracers.

%%%%%%%%%%%%%%%%%%%%%%%%%%%%%%%%%%%%%%%%%%%%%%%%%%%%%%%%%%%%%%%%
\section{Conclusion}
%%%%%%%%%%%%%%%%%%%%%%%%%%%%%%%%%%%%%%%%%%%%%%%%%%%%%%%%%%%%%%%%
In conclusion, we have presented a mechanism to control the plug formation in channels by turning on/off the self-propulsion. The working principle relies on the spontaneous formation of particle clusters preferentially near the walls.
The advantage of the method is the rapidity of the plug formation, even using very small densities of self-propelled particles.
This controlled clogging could be in practice achieved by exploiting the light-sensitivity of certain self-propelled particles, such as Janus colloids or genetically engineered \emph{E.~Coli} bacteria.
Furthermore, we expect the switching-mechanism to be more efficient in experimental devices than our simulated systems since Janus particles usually
make clustering at smaller densities with respect to numerical simulations~\cite{palacci2013living}.
In addition, a proper design of wall  geometries~\cite{nikola2016active, smallenburg2015swim, wysocki2015giant, wu2018transport, caprini2019transport} or the introduction of pillars in the bottleneck region~\cite{shi2020transport} can optimize the clogging
process taking advantage of enhanced trapping mechanisms~\cite{kaiser2012capture,mijalkov2013sorting,wu2018transport, kumar2019trapping}.

The clogging phase-diagram reported in Sec.~\ref{Sec:results} is obtained as a function of the density and the bottleneck width at fixed active force and, thus, persistence length, $v_0/D_r$.
%Except for very small values of the persistence length, at which clustering does not take place, 
We expect that the picture remains unchanged since the process is controlled by the ratio between the bottleneck width and the persistence length. 
The larger is the latter, than much favored is the clog formation.
For very small values of $v_0/D_r$, active systems behave as passive~\cite{fodor2016far, caprini2019activity} and, thus, clustering does not occur~\cite{digregorio2018full}.

Possible interesting improvements of this work towards a more realistic system would be:
i) studying the effects of a solvent through the inclusion of hydrodynamic interactions and ii) implementation of the flow. 
i) In our coarse-grained approach, the role of the solvent is only described as a thermal bath,
however, would be interesting to understand how the inclusion of the explicit solvent and the
consequent hydrodynamic interactions would change the phase diagram and the dynamical properties of the clogging process.
We expect that the switching mechanism is robust to the presence of hydrodynamics, indeed, it is known that the accumulation near obstacles and the clustering occur for both pushers and pullers microswimmers~\cite{malgaretti2017model, yoshinaga2017hydrodynamic}. 
%Nevertheless leaving the two-phase scenario (open/clogged states) unchanged. 
The presence of the explicit solvent opens new challenging
questions like the role of hydrodynamic pressure or osmotic pressure~\cite{rodenburg2017van, row2020reverse} in the clogged states.
ii) The explicit presence of a flow field pushing objects or debris in the channel is 
common in many microfluidic applications. This is not taken into account in this study, that is restricted to regimes of swim velocities where the fluid flow is negligible and does not consistently affect the active particle dynamics.
The addition of fluid flow and movable debris is a relevant issue that will be the
subject of future investigation to test the stability and resistance of the obstructions made by clustered active particles.

\appendix

%%%%%%%%%%%%%%%%%%%%%%%%%%%%%%%%%%%%%%%%%%%%%%%%%%%%%%%%%%%%%%%%%%%%%%%
\section{Geometrical set-up \label{Sec:Sec1}}
%%%%%%%%%%%%%%%%%%%%%%%%%%%%%%%%%%%%%%%%%%%%%%%%%%%%%%%%%%%%%%%%%%%%%%%
The container employed in the numerical study is formed by two lateral boxes of size $L \times H$ and a bottleneck of size $s \times w$, as shown in Fig.~\ref{fig:picture}~(a). 
The numerical set-up is obtained, by fixing $H=60$, $L=100$ and $s=50$ and varying $w$ in the range $[5,30]$.
The two lateral boxes, satisfying periodic boundaries conditions, are connected to each other by soft-walls whose shapes reproduce a narrow bottleneck.
The top bottleneck profile in the plane $x,y$ is described by a piece-wise 
function:
\begin{equation*}
k(x) = 
\dfrac{H-w}{2\pi} \arctan\left[K \left(x^2-\frac{s^2}{4}\right)\right]+\dfrac{H+w}{4}\,, % 0\le x \le s
\end{equation*}
for $0\le x \le s$ and $0$ elsewhere.
The bottom profile is a reflection around y-axis, $k(x)\to-k(x)$.
With this choice the bottleneck lies in the interval $(-s/2, s/2)$, while the left and right lateral boxes are placed in $(-L-s/2, -s/2)$ and $(s/2, s/2+L)$, respectively.
$K$ is the parameter which determines the sharpness of the corners formed by the lateral boxes and the bottleneck. 
Since the larger is $K$ the sharper are the corners, we chose $K=10$ in our numerical study.
The walls exert on the particles the force $\mathbf{F}_w$ directed along the normals with respect to the wall profiles. This direction is given by:
$$
\hat{\mathbf{n}}=\frac{(k'(x), -1)}{\sqrt{1+k'(x)}} \,,
$$
where the prime denotes the derivative with respect to $x$.
The amplitude of the force is the derivative of a harmonic potential truncated at its minimum
\begin{equation}
\label{eq:W(r)}
W(r)=A\frac{r^2}{2} \Theta(r) \,,
\end{equation}
where $A=10^3$ is the strength of the repulsive force chosen large enough to prevent the penetration of 
particles into the wall-regions. 

%%%%%%%%%%%%%%%%%%%%%%%%%%%%%%%%%%%%%%%%%%%%%%%%%%%%%%%%%
\section{Derivation of equation~(2)}\label{Sec:Sec2}
%%%%%%%%%%%%%%%%%%%%%%%%%%%%%%%%%%%%%%%%%%%%%%%%%%%%%%%%%
During the clogging, the density of the bottleneck region increases because of the particle flow from the lateral boxes towards the bottleneck region.
Self-propelled particles remain trapped in the bottleneck because of the interactions with the other particles which hinder their exit on the opposite side.
Because the self-propulsion forces change direction after a persistence time, $1/D_r$, we expect that the particles reaching the bottleneck are those contained in a square box of size given by the persistence length, $\lambda=v_0/D_r$.
We define $\nu_{\lambda}(s)$ as the rate of this process (number of particles per unit time).
Accordingly, the number of particles arriving at the bottleneck
in a time interval $[0,t]$ is 
\begin{equation}
Q(t)=\int_0^{t} ds \nu_{\lambda}(s) \approx \int ds D_r \lambda^2 \rho_{\lambda}(s) \mathcal{R} \,,
\end{equation}
where $\rho_{\lambda}(t)$ is the density in one of the two square regions of size $\lambda$ near the bottleneck and the factor $\mathcal{R}$ counts the fraction of particles able to reach the bottleneck region with velocity 
$v_0$ determined by the self-propulsion. 
This factor depends only on $\lambda$ and $w$ and will be estimated hereafter.

Since we took lateral boxes much larger than the bottleneck region ($L\gg s$), we can assume that $\rho_{\lambda}(t)$ remains nearly 
constant to its initial value, $\rho$.
In other words, the large size of the lateral boxes render negligible the loss of particles due to the flow into the bottleneck. 
With this approximation, we get
\begin{equation}
\label{eq:Q(t)_SM}
Q(t) \approx  \rho \,\mathcal{R}\lambda^2 D_r t \,.
\end{equation}
Now, we compute the clogging-time, $t_{\text{cl}}$, requiring that, in the clogged state, 
the maximal number of particles in the bottleneck is given by $N_m= \langle\rho_b\rangle w s$, where $\langle\rho_b\rangle$ is the maximal density admitted by the bottleneck region.
Neglecting the particles leaving this region, we get:
\begin{equation}
2 Q(t=t_w) + w s \rho=N_m \,.
\end{equation}
Using the explicit expressions for $Q(t)$ and $N_m$, we estimate $t_{cl}$
as 
\begin{equation}
\label{eq:tw_estimate_SM}
t_{\text{cl}} = \frac{w s (\langle\rho_b\rangle - \rho)}{2 D_r \mathcal{R} \lambda^2 \rho} \,.
\end{equation}
$t_{\text{cl}}$ assumes always positive values because $\langle\rho\rangle >\rho = \langle \rho(t=0)\rangle$, a condition which always occurs because of the particle accumulation at the walls.
We remark that the validity of this prediction requires the main hypothesis, $\rho_{\lambda}(t)\approx \rho_{\lambda}(0)=\rho$.

The simplest estimate of $\mathcal{R}$ is $\mathcal{R}=1/4$, assuming that particles move homogeneously in four directions, $\pm\hat{x}$, $\pm\hat{y}$. 
A more refined approximation consists in assuming that all the particles are placed in the middle of the square of size $\lambda$, at distance $\lambda/2$ from the center of the bottleneck.
In this case, the fraction of particles which can move towards the bottleneck can be obtained by geometrical arguments:
\begin{equation*}
\begin{aligned}
\mathcal{R} = \int^{\theta_{max}}_{-\theta_{max}} \frac{d\theta}{2\pi} =\frac{\theta_{max}}{\pi}=\frac{1}{\pi}\arctan{\frac{w}{\lambda}} \,.
\end{aligned}
\end{equation*}

%%%%%%%%%%%%%%%%%%%%%%%%%%%%%%%%%%%%%%%%%%%%%%%%%%%%%%%%%%%%%%%%%%%
\section{Shape of $\rho(t)$}\label{Sec:Sec3}

%--------------------------- FIG.5 --------------------------------------
\begin{figure*}[!t]
\centering
\includegraphics[width=1\linewidth,keepaspectratio]{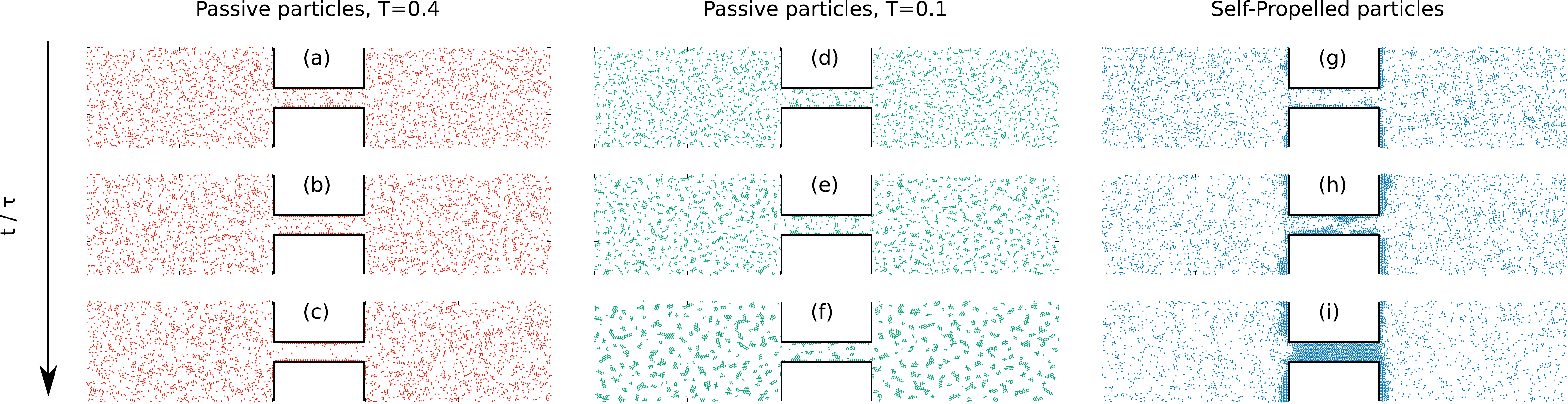}
\caption{\label{fig:SM}
Snapshot configurations obtained by simulations of particles in a channel of bottleneck width $w=10$ and density $\rho=0.2$ at different times: $t/\tau=1$ for panels~(a),~(d),~(g), $t/\tau=10$ for panels~(b),~(e),~(h) and $t/\tau=10^2$ for panels~(c),~(f),~(i).
Here, the time is normalized by the persistence time, $\tau$, of the active dynamics to make clear the comparison between passive and active cases.
Panels~(a)-(f) are obtained using attractive passive colloids following the dynamics~\eqref{eq:passive_dynamics}, at $T=0.1$ for~(a)-(c) and $T=0.4$ for~(d)-(f).
Finally, panels (g)-(i) refers to self-propelled particles, evolving with Eqs.~\eqref{eq:wholeABPdynamics}, using $v_0=25$ and $D_r=1$.
In the active and passive systems, simulations are run employing the same geometry, by setting $L=100$, $H=60$ and $l=50$.
}  
\end{figure*}
%--------------------------------------------------------------------------

%%%%%%%%%%%%%%%%%%%%%%%%%%%%%%%%%%%%%%%%%%%%%%%%%%%%%%%%%%%%%%%%%%%
Here, we derive the time behavior of $\rho_b(t)$.
Eq.~\eqref{eq:Q(t)_SM}, prescribes an non-physical unbounded growth of $\rho_b(t)$.
While such a simplified argument is sufficient to predict the clogging time $t_{\text{cl}}$, as discussed in the previous section, it cannot account for the behavior of $\rho_b(t)$ which, instead, stops increasing when the bottleneck is completely clogged.
To account for this saturation, we develop a differential equation to describe the time-evolution of $\rho_b(t)$. 
As already mentioned, the increase of $\rho_b(t)$ is due to the particles coming from the two lateral squares of size $\lambda\times\lambda$ (near the bottleneck).
Basically, in Eq.~\eqref{eq:tw_estimate_SM}, we are assuming that all the particles coming in the bottleneck remain trapped. 
However, the probability to remain trapped depends on the occupation degree of the bottleneck region (low occupation implies no trapping).
Thus, we expect that the probability of remaining trapped in the bottleneck is proportional to $\rho_b(t)$.
Additionally, when clusters are formed at the walls of the bottleneck, self-propelled particles behave as if the wall-width was $w_{eff} < w$. The shape of $w_{eff}$ depends on the density and can be estimated as: 
$$
w_{eff}=w \left(1-\frac{\rho_b(t)}{\langle\rho_b\rangle}\right) \,.
$$
As a result, we have:
$$
\dot{\rho}_b(t) \propto \rho_0 \mathcal{R} \lambda^2 D_r \rho_b(t) \propto \rho_b(t) \left(1-\frac{\rho_b(t)}{\langle\rho_b\rangle}\right) \,.
$$
with the initial condition $\rho_b(0)=\rho$.
The above differential equation admits a sigmoid solution which reads:
\begin{equation}
\label{eq:functionalform_rhob(t)}
\rho_b(t) = \frac{\langle\rho_b\rangle \rho}{\rho + \left(\langle\rho_b\rangle - \rho\right) e^{-t/\alpha}} \,, 
\end{equation}
where the characteristic time $\alpha$ is treated as a fitting parameter.

%%%%%%%%%%%%%%%%%%%%%%%%%%%%%%%%%%%%%%%%%%%%%%%%%%%%%%%%%%%%%
\section{The case of passive colloids}\label{Sec:Sec4}
%%%%%%%%%%%%%%%%%%%%%%%%%%%%%%%%%%%%%%%%%%%%%%%%%%%%%%%%%%%%%
The clogging-process shown in the channel geometry of Sec.~\ref{Sec:Sec1} works only employing suspensions of self-propelled particles, while a similar scenario cannot be observed using suspensions of pure repulsive passive colloids, at least waiting for a reasonable time.
In Sec.~\ref{Sec:model}, we have already shown that when the active force is turned off, the channel obstruction disappears because equilibrium repulsive colloids do not undergo clustering and the plug becomes unstable.
However, one can expect that, by introducing attractive interactions among particles and between particles and walls, a steady clogged state can be yet achieved. 
Its formation clearly will depend on the interplay between density and temperature.

To show that, even with attraction, passive colloids cannot clog the channel in reasonable times, we performed passive-particle simulations at density, $\rho$, in the geometrical setup used so far. 
Particles interact with the attractive version of the potential used for the active particles:
$$
U_{\text{len}}(r) = 4 \epsilon \left[\left(\frac{\sigma}{r}\right)^{12} - \left(\frac{\sigma}{r}\right)^6  \right] \,,
$$
with $\epsilon=\sigma=1$, like for the active system.
Additionally, also the walls of the bottleneck are attractive, and the dynamics is given by:
\begin{equation}
\label{eq:passive_dynamics}
\gamma\dot{\mathbf{x}}_i = - \nabla_i U_{tot}  + \mathbf{F}_i^w + \sqrt{2T} \boldsymbol{\eta}_i \,,
\end{equation}
where $\boldsymbol{\eta}$ is a white noise with zero average and unitary variance, and $T$ is the temperature of the thermal bath.
The potential $U_{tot}=\sum_{i<j} U_{\text{len}}(|\mathbf{x}_i -\mathbf{x}_j|)$ while $\mathbf{F}_i^w$ models the attractive force exerted by the walls.
The latter has the same form discussed in Sec.~\ref{Sec:Sec1}, with the only exception that it is not truncated at its minimum, $r=0$.
In practice, we replace $W(r)$ with $W_{att}(r)$, given by:
$$
W_{att}(r)=A \frac{r^2}{2} \Theta(r-\sigma) \,,
$$
which attracts particles at $r=0$ in a layer of width $\sigma$.

\begin{figure*}[!t]
\centering
\includegraphics[width=0.9\linewidth,keepaspectratio]{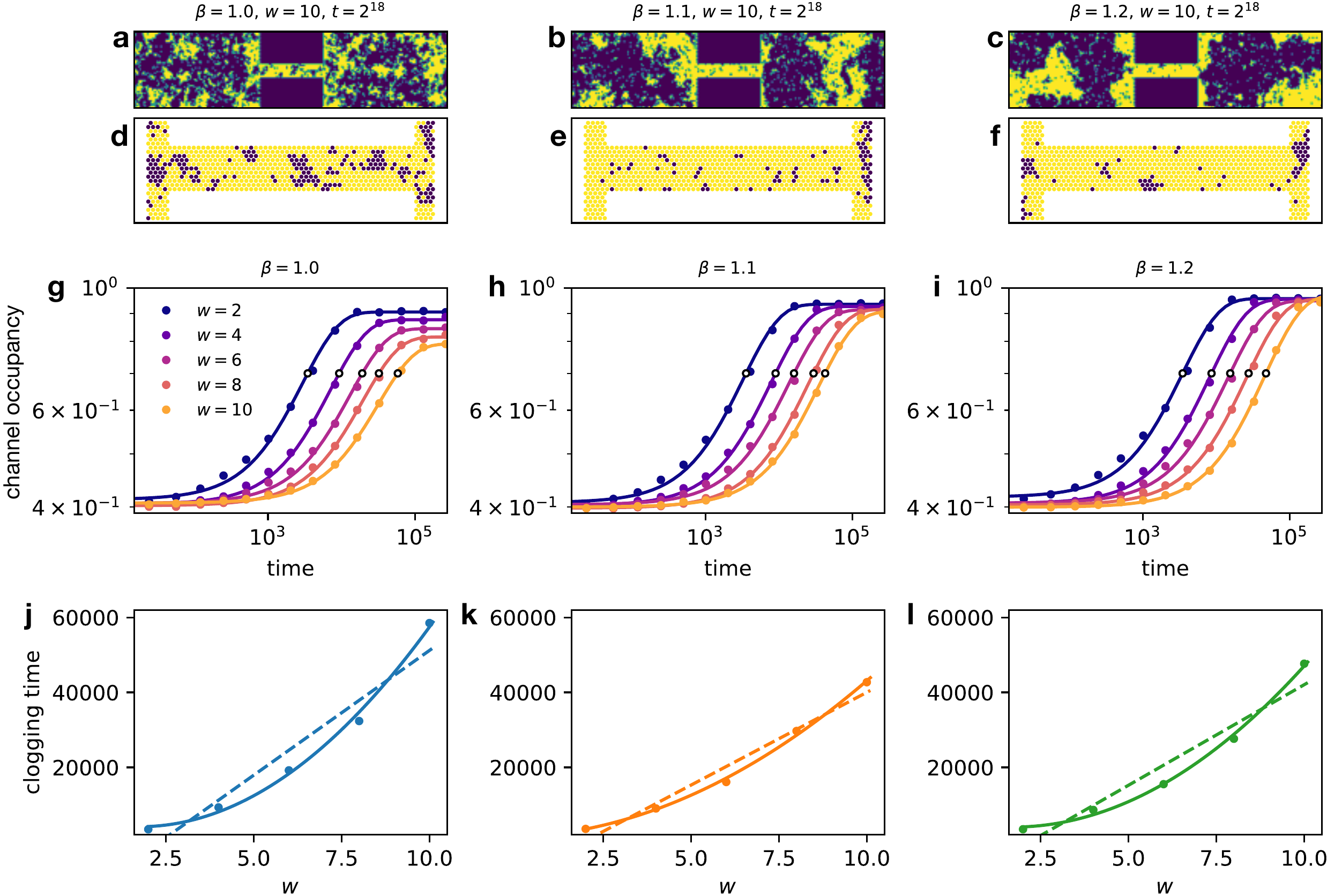}
\caption{\label{fig:latt}
Dynamics of a quenched lattice gas in a narrow channel.
Panels~(a),~(b) and~(c) show snapshots of the final configurations reached in MC simulations at fixed bottleneck width $w=10$ after a quench at different inverse temperatures: below $\beta_c$~(a), at $\beta_c$~(b) and above $\beta_c$~(c).
Panels~(d),~(e) and~(f) display the zoom of the bottleneck region of panels~(a),(b) and~(c), respectively. In panels~(g),~(h) and~(i), the average bottleneck occupancy is reported as a function of time for different $w$ and different $\beta$. 
The colored circles are the data from simulations, while the full lines are fits with exponential functions. The horizontally-aligned open circles mark the levels where the occupancy reaches the threshold value 0.7 (chosen as the clogging density). The abscissa of the interception between this threshold and the curve identifies the clogging time.
Panels~(j),~(k) and~(l) show the clogging time as a function of $w$ for the $\beta$ of panels~(g),~(h) and~(i), respectively. 
Here, the points are the clogging times from MC simulations, the dashed and solid lines are linear and quadratic fits of the data.}  
\end{figure*}
%-------------------------------------------------------------------

We run several simulations for different values of $T$ and $\rho$, fixing the bottleneck width $w=10$, for simplicity.
In this way, we consider a couple $(\rho, w)$ for which self-propelled particles clog the channel, as indicated by the clogging phase diagram, shown in Fig.~\ref{fig:phasediag}~(a).
Fig.~\ref{fig:SM} reports the spatial particle distribution in the channel at three successive times, for different cases: attractive passive-particles evolving with Eq.~\eqref{eq:passive_dynamics} for two different values of the temperature, $T=0.4$ (panels (a)-(c)) and $T=0.1$ (panels (d)-(f)), and,
for comparison, we show also the spatial distribution of self-propelled particle at the same times (panels (g)-(i)).
While the self-propelled particles clog the channel, passive attractive colloids are not able to perform a similar task in both cases, at least in the same time.
As shown by the first stage of their evolution, see Fig.~\ref{fig:SM}~(a) and~(c), passive particles form narrow layers near the walls of the bottleneck because of the particles-wall short-range attraction.
The two temperature values are chosen to show the two different scenarios occurring by varying the temperature: for low value of $T$, particles in the lateral boxes form many small clusters due to the attractive components of the interaction, while, for larger $T$, clusters in the lateral boxes disappear being destroyed by thermal fluctuations.
Even in the former case, particles starting from the homogeneous distribution attain a metastable state with many small clusters in the lateral boxes which cannot easily diffuse towards the bottleneck.
We are not able to state whether the passive system could eventually reach the clogged configuration, we can only state that a clogged process will require a time much longer than the time taken by the active counterpart. 
In fact, passive colloids can approach the bottleneck only by diffusion, that is a process intrinsically too slow to compete with the self-propelling dynamics.

As a conclusion, passive systems cannot be really useful to develop a clogging mechanism that can be used as a relatively fast switch.

%%%%%%%%%%%%%%%%%%%%%%%%%%%%%%%%%%%%%%%%%%%%%%%%%%%%%%%%%%%%%%%%%%%
\section{Lattice gas modeling of channel clogging \label{Sec:Sec5}}

%%%%%%%%%%%%%%%%%%%%%%%%%%%%%%%%%%%%%%%%%%%%%%%%%%%%%%%%%%%%%%%%%%%
Despite the system of passive particles interacting through Lennard-Jones potentials do not show clogged states in reasonable times, studying the dynamical features distinguishing thermal and active clogging could still represent an interesting issue.
To shed light on this point, we consider a lattice gas on a triangular grid over the channel geometry employed so far.
By imposing periodic boundary conditions every site has $6$ neighbors and the total Hamiltonian is given by
\begin{equation}
\label{eq:dynamics_MC}
H_\mathrm{lg} = - J \sum_{\langle i,j\rangle} n_i n_j \,,
\end{equation}
where $J$ is the coupling constant, set to $1$ for convenience, and $n_i$ is the occupancy of the $i$-th site which assumes the values $0$ or $1$. 
We have implemented simulations of a large system composed by $N=N_x \times N_y$ sites (with $N_y=64$, $N_x=64\times4$ and $N=16384$). 
These sites are enclosed in a rectangular box of size $(0, L) \times (0, H)$ 
with $L = a\,N_x$ and $H = a \sqrt{3} N_y/2$, where $a=1$ is the lattice spacing.
The simulations conserve the total occupancy (i.e. $\sum_i n_i = \mathrm{const}$) by using the Kawasaki dynamics in which a site can exchange its occupancy only with its neighboring sites~\cite{bovier2015kawasaki}. 
After this switch, a standard Monte Carlo (MC) Metropolis rule is applied and the new configuration is accepted or rejected according to the energy change.
All the numerical results are obtained starting from random configurations (i.e. at infinite temperature) with fixed total occupancy $\sum_i n_i = 6553$, corresponding to an average occupancy $\frac{1}{N}\sum_i n_i \approx 0.4$ which is below the critical one.
To simulate the presence of an attractive channel wall, we freeze to $n_i=1$, the sites placed at the positions $(x,y)$ such that $|y-H/2|>w/2$ and $|x-L|<s/2$, that are never updated in the MC simulation. 
In one MC step, we pick $N$ random sites and we attempt to switch the occupancy of each site with the occupancy of one of its neighbors. 

The transformation $n_i = (1+\sigma_i)/2$ maps the model~\eqref{eq:dynamics_MC} onto the Ising model on the triangular lattice, having critical temperature $T_c=4/\ln 3 \approx 3.641$ (for $J=1$ and $k_B=1$)~\cite{zhi2009critical}.
As a consequence, the critical temperature of the lattice gas model turns to be $T_c = (\ln 3)^{-1}$ (i.e. an inverse critical temperature $\beta_c = {T_c}^{-1} \approx 1.099$) while the critical average occupancy is $n_c = 1/2$.
Using this information, we can simulate the triangular lattice gas undergoing condensation in a channel geometry analogous to that of the active system, by varying the inverse temperature around $\beta_c$.
By quenching this system slightly above $\beta_c$ at $\beta_c$ and below $\beta_c$, after $2^{18}$ MC steps, we observe that the sites in the channel are preferentially occupied (i.e. the channel is clogged). 
This is shown in Fig.~\ref{fig:latt} where the occupied and empty sites are drawn in yellow and in violet, respectively, for a quenching: below (a), 
at (b), and above (c) $\beta_c$, respectively (note that, for graphical reasons, the sites in the channel walls are colored in violet instead of being yellow). 
The clustering in the channel is significantly more compact for $\beta \geq \beta_c$ as shown by the zoomed channel configuration reported in Figs.~\ref{fig:latt}~(d),~(e) and~(f) (note that, here, the sites of the channel walls are not plotted). 
%------------------------------------------------------------------ 
\begin{figure}
\centering
\includegraphics[width=0.95\columnwidth,keepaspectratio]
{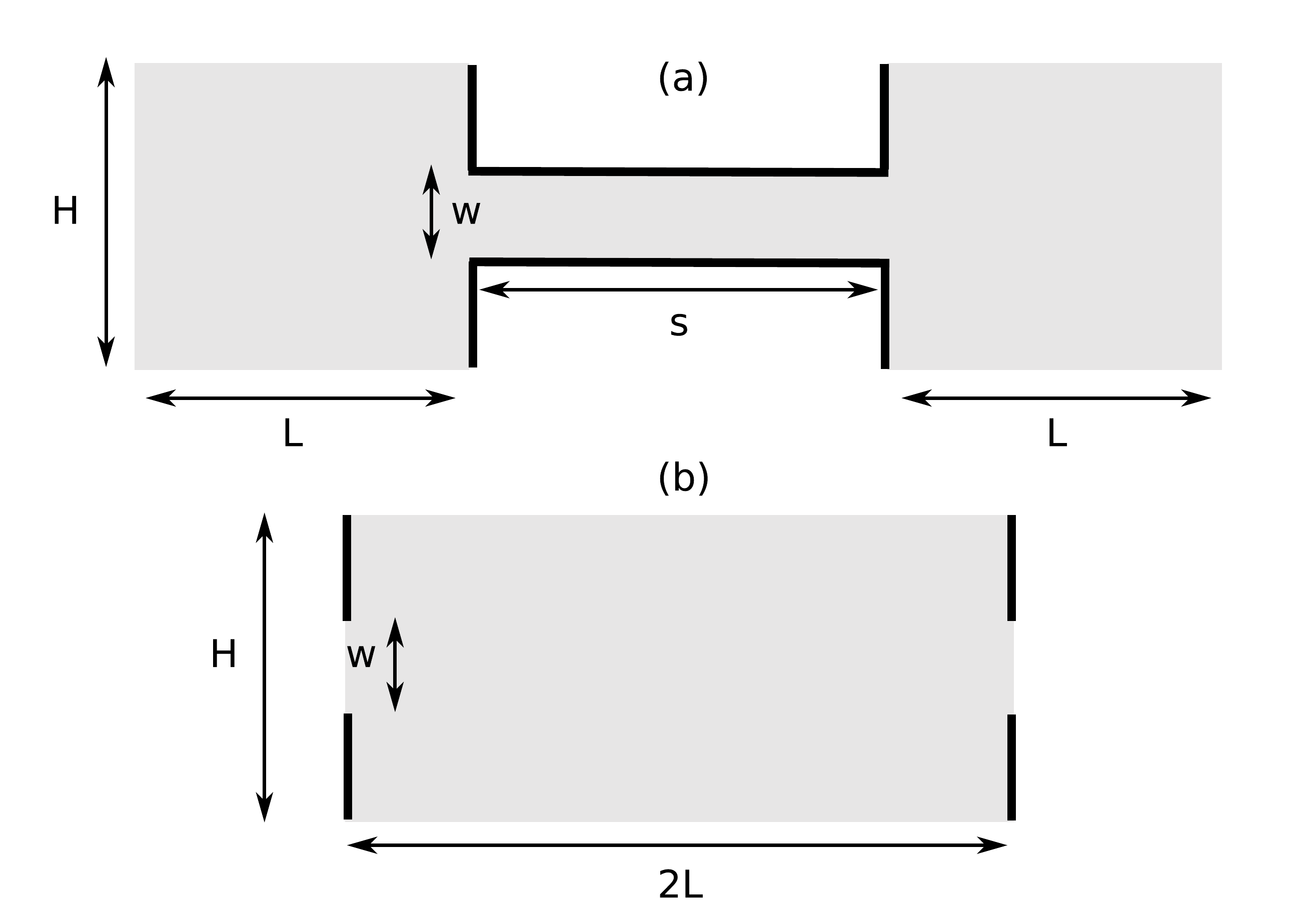}
\caption{\label{fig:box} Panel~(a): sketch of the channel employed in the numerical study. 
Panel~(b): rearrangement of the channel geometry to map the bottleneck filling 
of the original problem into the emptying process of the bulk reservoirs. 
In both panels, solid black lines represent repulsive walls, while their absence denotes periodic boundary conditions, except for the two segments of panel~(b) 
of width $w$, which are absorbing.
}
\end{figure}
%------------------------------------------------------------------ 
We monitor the clogging process by measuring the average occupancy in the bottleneck, 
$$
n_\mathrm{ch} = \dfrac{1}{N_\mathrm{ch}}\sum_i'n_i \,,
$$
as a function of time, where the prime indicates that the sum includes only the $N_\mathrm{ch}$ sites within the bottleneck region. 
The behavior of $n_\mathrm{ch}$ versus time is plotted in Figs.~\ref{fig:latt}~(g),~(h) and~(i) for various values of the channel width $w$ and quench temperature $\beta$ (the data points represent the average result of $128$ independent runs). 
It is clear that, during the channel clogging, the occupancy grows from the initial value $0.4$ to values close to the full occupancy. 
However, for quenches below $\beta_c$, we find that the final occupancy depends sensitively on the channel width (see Fig.~\ref{fig:latt}~(g)). 
It is also evident that larger channels need more time to be clogged by the lattice gas (the points progressively shift towards higher times as $w$ increases at fixed $\beta$). 
We also note that, at least for large $w$ and large times, the data points are always well-fitted by exponential functions.

We define the clogging time as the time where $n_\mathrm{ch}$ reaches the (arbitrarily) threshold $n_\mathrm{ch}=0.7$ (open circles in Figs.~\ref{fig:latt}~(g),~(h) and~(i)). 
In Figs.~\ref{fig:latt}~(j),~(k) and~(l), we report the clogging time as a function of $w$ for quenches above $\beta_c$, at $\beta_c$ and below $\beta_c$, respectively. 
From the linear and quadratic fits of the data in these figures, we conclude that the clogging time grows faster than linear with the channel width $w$.
This quadratic behavior is in contrast with the linear $w$-scaling observed for self-propelled particles, reported in Fig.~\ref{fig:timeandtraj}, supporting again the statement that passive clogging is less efficient than the active one.

%%%%%%%%%%%%%%%%%%%%%%%%%%%%%%%%%%%%%%%%%%%%%%%%%%%%%%%%%%%%%%%%%%%%%
\section{Diffusion Model for passive system}\label{Sec:Sec6}
%%%%%%%%%%%%%%%%%%%%%%%%%%%%%%%%%%%%%%%%%%%%%%%%%%%%%%%%%%%%%%%%%%%%%
%---------------------------- FIG.2 ----------------------------------
\begin{figure}[t]
\centering
\includegraphics[width=0.9\columnwidth,keepaspectratio]
{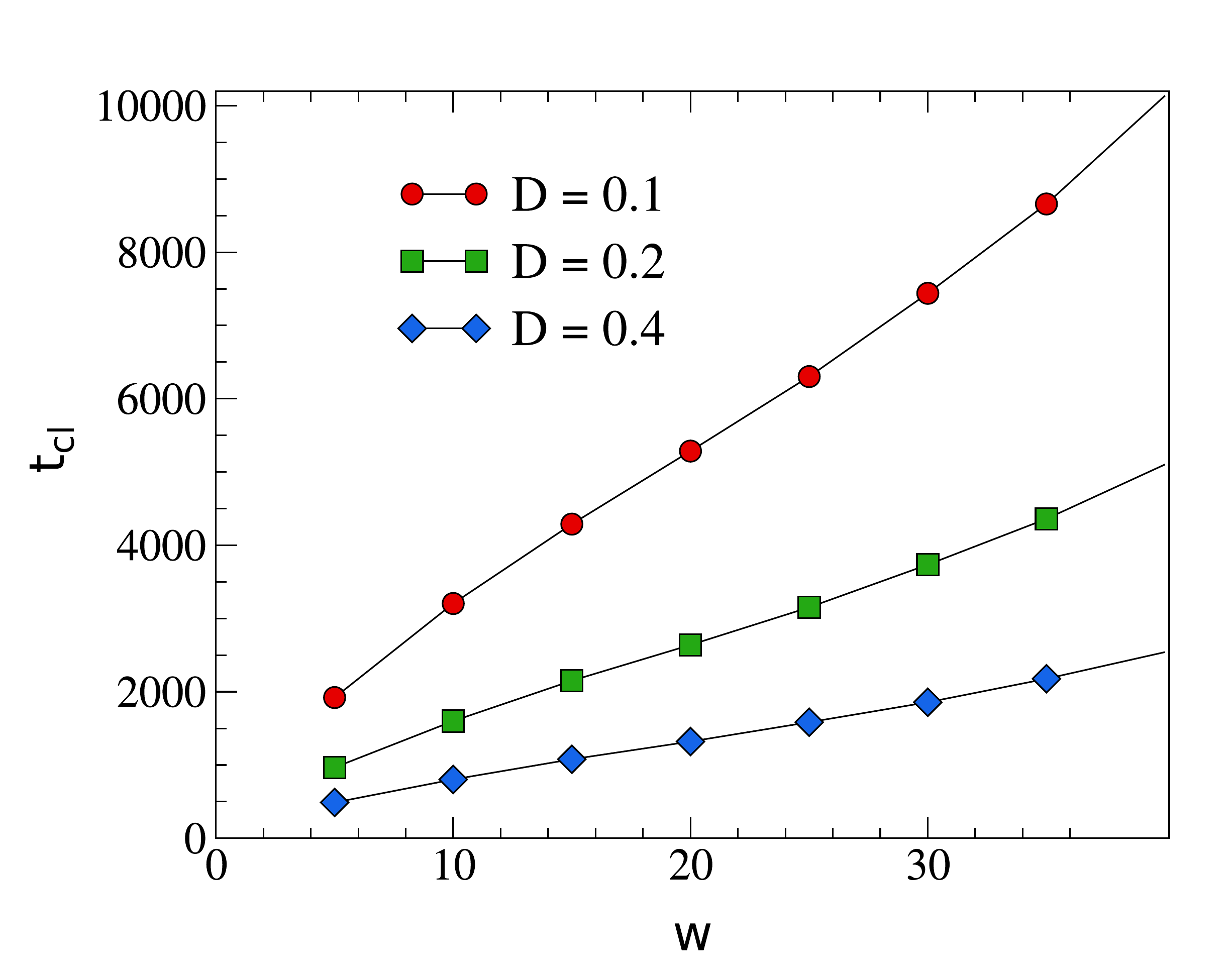}
\caption{\label{fig:scaling} Clogging time, $t_{\text{cl}}$, as a function of the bottleneck width, $w$, for three values of the diffusion 
coefficient $D$, in Eq.~\eqref{eq:diffusiveequation}, and density $\rho= 0.4$. 
Each point is the result of an average over $M=100$ independent initial
configurations corresponding to a homogeneous particle distribution
in the box $2L\times H$.}
\end{figure}
%----------------------------------------------------------------------
To explain why the passive clogging has not been observed for passive systems,  
we study the diffusive dynamics of an assembly of non-interacting particles and mimic the clogging process studied in this paper by using a suitable restricted geometry.
This test provides a lower bound for the clogging time obtained with a passive system with attractive interactions because attraction reduces the effective diffusion of the single-particle.

 Since we are interested in the emptying dynamics of the lateral reservoirs, it is useful to shift the system of $L+s/2$ along the $x$-axis.
In this way, due to the periodic boundary conditions along $x$,  
the system appears as a single box with two small lateral apertures of width $w$ (trace of the bottleneck presence) see Fig.~\ref{fig:box}.
In what follows, we refer as box to denote the reservoirs of the original problem.
The emptying process of the box (responsible for the clogging) is simulated by replacing the bottleneck by two symmetric absorbing boundaries of width $w$ placed at $\pm L$. 
Therefore, the absorbed particles are virtually those clustered in the bottleneck region.
In this way, we are neglecting any coarsening process occurring in the box assuming that any particle absorbed into the bottleneck cannot come back.

More specifically, the box initially contains an ensemble of $2LH\rho$ particles uniformly distributed with density $\rho$, evolving according to a diffusive dynamics
\begin{equation}
\dot{\mathbf{x}}= \sqrt{2\gamma D} \boldsymbol{\zeta} \,,
\label{eq:diffusiveequation}
\end{equation}
where $\gamma$ is the friction due to the solvent and $D$ the diffusion coefficient. The term $\boldsymbol{\zeta}$ is a white noise with zero average and unit variance.
The parameters are chosen to reproduce the experimental conditions corresponding to room temperature.
To account for the clogging phenomenology, we choose mixed boundary conditions along the box perimeter, i.e. they are periodic on the two edges of size $2L$, and 
reflecting on the $H$ sides, except for the two apertures (that mimic the bottleneck) of width $w$, which are absorbing (as shown in Fig.~\ref{fig:box}~(b)).

We run simulation of the escaping process from the box to get an estimate of the clogging time, $t_{\text{cl}}$. Since the absorbed particles correspond to the particles migrating to the bottleneck of the original problem, $t_{\text{cl}}$ will be the first time
at which the number of absorbed particles equals the maximal number of particles contained in the bottleneck (roughly, at packing density $\rho_p=1.2$).
Fig.~\ref{fig:scaling} displays the clogging time, $t_{\text{cl}}$, as a function of the bottleneck width, $w$, showing that, even in the non-interacting passive case, the time needed to clog the bottleneck is at least two or three orders of magnitudes longer than the time required by the active clogging. 
This explains why the passive Brownian system with attractive interactions (Sec.~\ref{Sec:Sec4}) does not exhibit the clogging process in reasonable times.

\bibliographystyle{apsrev4-1}

\bibliography{clogging.bib}

\end{document}